\documentclass[aps,reprint,amssymb,superscriptaddress]{revtex4-2}  
\usepackage{times,units,braket}
\usepackage{graphicx}  
\usepackage{graphics}
\usepackage{dcolumn}   
\usepackage{bm}        
\usepackage{amssymb}   
\usepackage{amsmath,upgreek}
\usepackage{natbib}
\usepackage[colorlinks,citecolor=blue]{hyperref}
\usepackage{color}
\usepackage{amssymb}
\usepackage{changepage}
\usepackage{float}
\usepackage[utf8]{inputenc}
\begin{document}

\title{Tunneling photons pose no challenge to Bohmian mechanics}
\author{Yun-Fei Wang}
\author{Xiao-Yu Wang}
\author{Hui Wang}
\affiliation{%
    Hefei National Research Center for Physical Sciences at the Microscale and School of Physical Sciences, University of Science and Technology of China, Hefei 230026, China%
}
\affiliation{%
    Shanghai Research Center for Quantum Science and CAS Center for Excellence in Quantum Information and Quantum Physics, University of Science and Technology of China, Shanghai 201315, China%
}
\affiliation{%
    Hefei National Laboratory, University of Science and Technology of China, Hefei 230088, China%
}

\date{\today}

\begin{abstract}
   Very recently, Sharoglazova et al. performed an experiment measuring the energy-velocity relationship and Bohmian velocity in coupled waveguides. Their data show a discrepancy between the semi-classical `speed' $v=\sqrt{2|\Delta|/m}$ and Bohmian velocity $\vec{v}_s$ for $\Delta<-\hbar J_0$, leading them to claim a challenge to Bohmian mechanics. Here, we definitively demonstrate this experiment poses no challenge to Bohmian mechanics. First, $v$ and $\vec{v}_s$ represent fundamentally distinct physical quantities---comparing a scalar and a vector is physically unjustified and cannot adjudicate between Copenhagen and Bohmian interpretations. Second, we rigorously show that both interpretations predict identical photon tunneling dynamics in coupled waveguides. 
\end{abstract}

\maketitle

Bohmian mechanics (BM) is an alternative interpretation of quantum mechanics, firstly proposed by de Broglie in 1927~\cite{De27} and further developed by Bohm in 1952 \cite{Bohm52}. It is widely believed that BM has the same prediction on experimental outcomes with textbook Copenhagen interpretation because they obey the same fundamental Schr\"{o}dinger equation in the non-relativistic regime. 

However, a recent experiment by Sharoglazova et al. \cite{S25} reports an apparent discrepancy between measured energy-velocity relationships and Bohmian predictions in specific regimes. The authors contend this mismatch ``casts doubt on whether the guiding equation accurately captures the temporal dynamics of the scattering process". 

In their experiment, the authors employ a high-finesse microcavity containing a dye medium. Within this cavity, the engineered height difference is implemented for the purpose of measuring speed of photons. First, a linear ramp in the main waveguide is designed to control the initial kinetic energy of photons by changing the position along this ramp at which the dye is optically pumped. Subsequently, the emitted photons impinge a constant step potential $V_0$, where another auxiliary waveguide starts. This auxiliary waveguide evanescently couples to the main waveguide with a fixed coupling strength $J_0$. This system is modeled by two coupled Schr\"{o}dinger equations
\begin{equation}
\begin{aligned}
i\hbar \frac{\partial}{\partial t}\psi_m&=-\frac{\hbar^2}{m}\frac{d^2 \psi_m}{dx^2}+V_0 \psi_m+\hbar J_0 (\psi_a - \psi_m), \\
i\hbar \frac{\partial}{\partial t}\psi_a&=-\frac{\hbar^2}{m}\frac{d^2 \psi_a}{dx^2}+V_0 \psi_a+\hbar J_0 (\psi_m - \psi_a),
\label{s equation}
\end{aligned}
\end{equation}
where $m$ is the effective mass of photons considering the longitudinal resonance energy $E_z$ using Einstein energy-mass equivalence $E_z=m \tilde{c}^2$ ($\tilde{c}$ denotes the photon speed in the medium), $\psi_m$ and $\psi_a$ are wavefunctions in the main and auxiliary waveguides, respectively.

\begin{figure}
    \centering
    \includegraphics[width=1\linewidth]{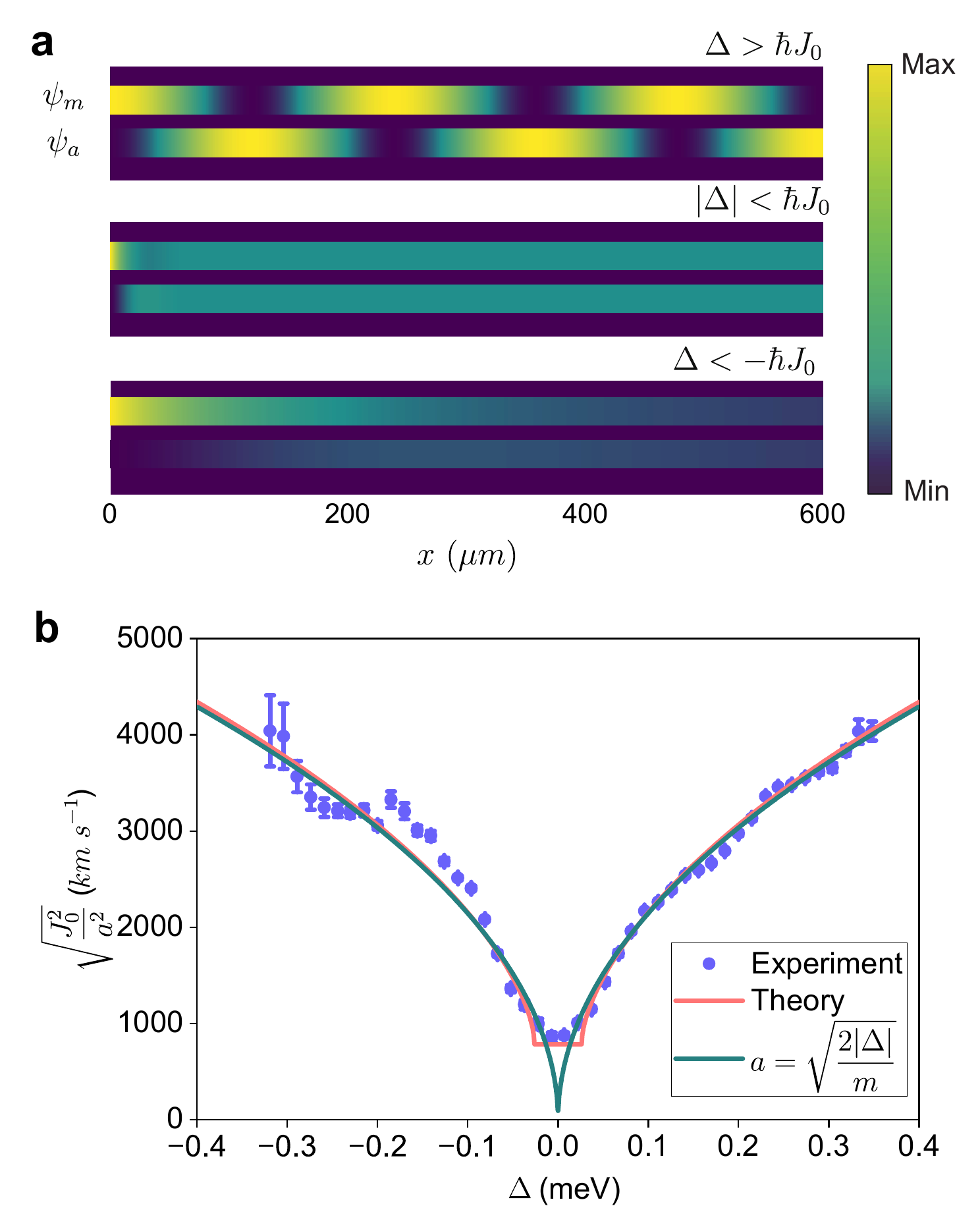}
    \caption{\textbf{Quantum tunneling between two coupled wavegiudes based on Schr\"{o}dinger equation}. \textbf{a,} Calculated wavefunctions $\psi_m$ and $\psi_a$ along $x$ direction. Three conditions are considered here with different $\Delta$. The upper panel shows two transmission modes, the middle panel presents one transmission and evanescent modes, and the lower panel displays two evanescent modes. \textbf{b,} Purple dots denotes and cyan line are experimental determined $v$ and fitted results in Ref.~\cite{S25}, respectively. Red line is our theoretical fitting to the experimental results in Ref.~\cite{S25}.
    }
    \label{fig1}
\end{figure}

The authors perform two distinct `velocity' measurements. The first method estimates a semi-classical speed $v$ through
\begin{equation}
    \rho_a=\left(\frac{J_0 x}{v}\right)^2,
    \label{velocity1}
\end{equation}
where $\rho_a=|\psi_a|^2/(|\psi_a|^2+|\psi_m|^2)$. Thus $v$ can be determined by recording the populations $|\psi_a|^2$ and $|\psi_m|^2$ in the auxiliary and main waveguides at different $x$ positions, respectively. Noting that this is not velocity measurement since the direction of motion can not be determined. The authors say that this speed ``represents the local speed of particles directly behind
the potential step", but the direction of this speed is still unclear. The second velocity is determined using guiding equation $\vec{v}_s=\frac{\hbar}{m}\nabla S$ of BM through ``interfere quantum states in the step potential with their spatially mirrored images to
detect phase gradients". Based on the authors' claim of $\vec{v}_s=0$ for negative kinetic energy case, we believe that this $\vec{v}_s$ should along the $x$ direction (see Supplementary information). 

\begin{figure*}
    \centering
    \includegraphics[width=1\linewidth]{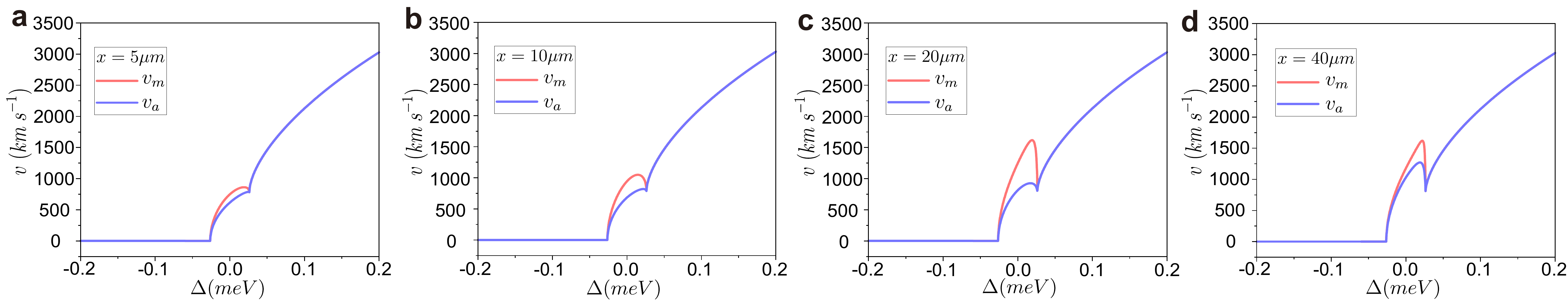}
    \caption{\textbf{Quantum tunnelling between two coupled waveguides based on Bohmian mechanics}. Four group of velocities at $x$=5 $\mu$m (\textbf{a}), 10 $\mu$m (\textbf{b}), 20 $\mu$m (\textbf{c}) and 40 $\mu$m (\textbf{d}) are displayed. If $\Delta < -\hbar J_0$, both $v_m$ and $v_a$ along $x$ are zero. In the regime of $|\Delta| < \hbar J_0$, $v_m$ and $v_a$ are distinguished because of the different quantum potential. For $\Delta > \hbar J_0$, $v_m$ and $v_a$ are identical. 
    }
    \label{fig2}
\end{figure*}

The discrepancy of these two measured speeds happens when $\Delta < -\hbar J_0$, where $\Delta=E - V_0 + \hbar J_0$. The observed $v$ is nonzero and is proportional to $\sqrt{\frac{2|\Delta|}{m}}$, while measured $\vec{v}_s$ is very close to zero. Therefore, the authors claim that this experiment raises concern on the correctness of BM. 

Here we reanalyze the physical process in this experiment and demonstrate that it poses no challenge to BM. First, the two measured speeds represent fundamentally distinct physical quantities: $v$ is a scalar, which quantifies the magnitudes of velocity while the direction is unclear; $\vec{v}_s$ is the velocity defined by the continuity equation associated with Schr\"{o}dinger equation, thus it's independent with the Copenhagen interpretation or the BM. Comparing a scalar and a vector is physically unjustified and cannot serve as evidence favoring the Copenhagen interpretation over BM. Second, we theoretically prove that both the widely-held Copenhagen interpretation and BM yield identical predictions for photon tunneling in coupled waveguides, thereby resolving the contradiction raised by Ref.~\cite{S25}.

We solve the equation~\ref{s equation} to obtain the solution 
\begin{equation}
\begin{aligned}
\psi_m& \propto (e^{ik_+ x}+e^{ik_- x}) \\
\psi_a& \propto (e^{ik_+ x}-e^{ik_- x}),
\label{solution}
\end{aligned}
\end{equation}
assuming no back-propagating modes and symmetric distribution of $k_+$ and $k_-$ across both waveguides (for simplicity), where $k_\pm=\sqrt{\frac{2m(\Delta \mp \hbar J_0)}{\hbar^2}}$. The dynamical regimes are characterized by $\Delta$. When $\Delta > \hbar J_0$, both $k_+$ and $k_-$ are real values, indicating there are two transmission modes in waveguide. When $|\Delta| < \hbar J_0$, $k_+$ is imaginary, representing an evanescent mode, while $k_-$ is real which still a transmission mode. When $\Delta < -\hbar J_0$, only evanescent modes exist in the waveguide, thus photon population exponentially decays. Numerical results in Fig.~\ref{fig1}a confirm these predictions and align with experimental observations from Ref.~\cite{S25}.

Note that our solution differs from Ref.~\cite{S25} and better describes their experimental results. For sufficiently small $x$, a Taylor expansion of $\psi_m$ and $\psi_a$ yields
\begin{equation}
\begin{aligned}
\rho_a=&\frac{2m|c_0|^2}{\hbar^2}(|\Delta+\hbar J_0| - 2\sqrt{\text{Max}\{\Delta^2-(\hbar J_0)^2,0 \}} \\
&+ |\Delta-\hbar J_0|)^2x^2+O(x^3).
\label{population}
\end{aligned}
\end{equation}
Similar to Ref.~\cite{S25}, $\rho_a$ is quadratic in $x$ but with a distinct coefficient. Fitting their defined speed $v$ (Fig.~\ref{fig1}b) shows that our result (red line) exactly reproduces the experimental data, especially in the region $|\Delta| < \hbar J_0$ where a plateau appears. However, the results in Ref.~\cite{S25} fail to describe this plateau.

The authors in Ref.~\cite{S25} model $\rho_a$ as $\rho_a = (J_0 x/v)^2$. Though they emphasize this speed $v$ ``represent the  the local speed of particles directly behind the potential step", we still have concerns whether this is acceptable. Even if accepted, $v$ cannot be directly compared to the Bohmian velocity ($v_s$) defined by the guiding equation. First, the direction of $v$ is ambiguous---does it include a $y$-component? In contrast, $v_s$ is strictly along the $x$-direction (see Supplementary information). Second, $v_s$ inherently satisfies the continuity equation derived from the Schrödinger equation, whereas $v$ does not. Any speed rigorously defined within the Schrödinger formalism must be equivalent to $v_s$ to ensure consistency with probability conservation.

Next, we derive the Bohmian velocity $v_s$ from continuity equation:
$v_s(x)=\frac{\hbar}{m}\text{Im}\frac{\psi^*\partial_x\psi}{\psi^*\psi}$. Using the solution in equation~\ref{solution}, $v_{s,m}(x)$ and $v_{s,a}(x)$ can be determined. Fig.~\ref{fig2} shows these calculated velocities at four position ($x=5, 10, 20, 40$ $\mu$m). In the region of $\Delta < -\hbar J_0$, all velocities vanish. This is expected: the wavefunction decays exponentially (purely evanescent), indicating photons are stationary along $x$. Crucially, this does not preclude tunneling---finite probabilities $|\psi_a|^2$ and $|\psi_m|^2$ persist in the waveguides, enabling tunneling despite zero net flow.

To definitively demonstrate this experiment poses no challenge to BM, we prove BM yields identical photon populations $\rho_a$ to the Schrödinger solution (equation~\ref{population}; full derivation in Supplementary information). First, substituting the wavefunction in equation (11) in SM into the Schrödinger equations yields new quantum Hamilton-Jacobi equations and continuity equations. The continuity equations contain a tunneling current term $j_0 = 2 J_0 R_a R_m \text{sin}(S_a - S_m)$, which accounts for inter-waveguide quantum tunneling while preserving probability conservation. Solving the reduced continuity equation $\partial_x(\rho_{a,B} v_{a, x})+j_0=0$ with BM-derived $v_{a,x}$ and $j_0$ gives $\rho_{a,B}$, which is presented in equation (18) in Supplementary information. Subsequent analysis confirms exact agreement with the Schrödinger-predicted $\rho_a$ (equation~\ref{population}). This result indicates that (i) the Bohmian velocity $v_{a,x}$ generates the same $\rho_a$ as standard Schr\"{o}dinger equation, and (ii) both interpretations predict identical tunneling dynamics, refuting claims in Ref.~\cite{S25}.

In conclusion, the experiment in Ref.~\cite{S25} poses no challenge to BM. The mistakes in Ref.~\cite{S25} primarily stem from misinterpreting the semi-classical speed $v$ (equation~\ref{velocity1}) and Bohmian velocity $v_s$ as comparable quantities. We point out that these represent fundamentally distinct physical entities, their discrepancy cannot validate claims against BM. Furthermore, solving for $\rho_a$ through both Schr\"{o}dinger equation and continuity equation in BM yields identical results. This confirms no interpretational distinction exists between Copenhagen and BM for quantum tunneling phenomena. In other words, any velocity defined from $\rho_a$ must be identical in both frameworks, meanwhile satisfy continuity equation. Last but not least, $v_s=0$ in the region of $\Delta < -\hbar J_0$ is physically expected. From the perspective of BM, photons are indeed at rest in the step potential along $x$ direction, but by no means it imply that quantum tunneling does not occur. In fact, non-vanishing probability of evanescent wavefunction still allows quantum tunneling to come up.

\clearpage

\begin{titlepage}
  \begin{center}
    {\large \textbf{Supplementary information for\\ Tunneling photons pose no challenge to Bohmian mechanics}}\\[1cm]
  \end{center}
\end{titlepage}

\noindent{\textbf{Deriving the $\psi_m$, $\psi_a$ solution from Schr\"{o}dinger equation}}

The two coupled Schr\"{o}dinger equations in the main text satisfy:
\begin{equation}
\begin{aligned}
i\hbar\partial_t(\psi_m+\psi_a)=&-\frac{\hbar^2}{2m}\frac{d^2(\psi_m+\psi_a)}{dx^2}+V_0(\psi_m+\psi_a),\\
i\hbar\partial_t(\psi_m-\psi_a)=&-\frac{\hbar^2}{2m}\frac{d^2(\psi_m-\psi_a)}{dx^2}\\
&+V_0(\psi_m-\psi_a)-2\hbar J_0(\psi_m-\psi_a).
\end{aligned}
\end{equation}
We get two solutions:
\begin{equation}
\begin{aligned}
\psi_m+\psi_a&=A_+ e^{ik_+ x}+B_+e^{-ik_+ x},\\
\psi_m-\psi_a&=A_- e^{ik_- x}+B_-e^{-k_- x},
\end{aligned}
\end{equation}
where
\begin{equation}
\begin{aligned}
k_+&=\sqrt{\frac{2m(E-V_0)}{\hbar^2}}=\sqrt{\frac{2m(\Delta-\hbar J_0)}{\hbar^2}},\\
k_-&=\sqrt{\frac{2m(E-V_0+2\hbar J_0)}{\hbar^2}}=\sqrt{\frac{2m(\Delta+\hbar J_0)}{\hbar^2}},
\end{aligned}
\end{equation}
$\Delta=E-V_0+\hbar J_0 $, and $A_i, B_i$ is the amplitude of each transmission mode.

Subsequently, the solution of $\psi_m$ and $\psi_a$ is:
\begin{equation}
\begin{aligned}
\psi_m&=\frac{1}{2}(A_+ e^{ik_+ x}+A_- e^{ik_- x}+B_+e^{-ik_+ x}+B_-e^{-ik_- x}),\\
\psi_a&=\frac{1}{2}(A_+ e^{ik_+ x}-A_- e^{ik_- x}+B_+e^{-ik_+ x}-B_-e^{-ik_- x}).
\end{aligned}
\end{equation}
According to the initial conditions that the initial auxiliary mode is zero and there is no back propagation mode, we have the following:
\begin{equation}
\begin{aligned}
A_+&=c_0,\\
A_-&=c_0,\\
B_+&=0,\\
B_-&=0.
\end{aligned}
\end{equation}
Finally two transmitted wavefunctions are given by:
\begin{equation}
\begin{aligned}
\psi_m&=\frac{c_0}{2}(e^{ik_+ x}+e^{ik_- x}), \\
\psi_a&=\frac{c_0}{2}(e^{ik_+ x}-e^{ik_- x}).
\label{wavefunction1}
\end{aligned}
\end{equation}
This result indicates that two modes ($k_+$ and $k_-$) exist in two coupled waveguides with different energies.
\\

\noindent{\textbf{Analyzing the behavior of $\rho_a$ when $x$ is small}}

Based on equation~\ref{wavefunction1}, the population of auxiliary mode $\rho_a=|\psi_a|^2$ (it can be checked that the normalization method in Ref.~\cite{S25} will not change the Taylor expansion behavior at $x=0$) can be derived as:
\begin{equation}
\begin{aligned}  
\rho_a=&\frac{|c_0|^2}{4}(e^{i(k_+-k_+^*)x}-e^{i(k_--k_+^*)x}\\
&-e^{i(k_+-k_-^*)x}+e^{i(k_--k_-^*)x}).
\end{aligned}
\end{equation}
When $x$ is small, we can expand $\rho_a$ at $x=0$:
\begin{equation}
\begin{aligned}
\rho_a=&-\frac{|c_0|^2}{8}((k_+-k_+^*)^2+(k_--k_+^*)^2\\
&+(k_+-k_-^*)^2+(k_--k_-^*)^2)x^2+O(x^3)\\
=&\frac{m|c_0|^2}{2\hbar^2}(|\Delta+\hbar J_0|-2\sqrt{\text{Max}\{\Delta^2-(\hbar J_0)^2,0\}}\\
&+|\Delta-\hbar J_0|)x^2+O(x^3).
\end{aligned}
\end{equation}
Similar to Ref.~\cite{S25}, $\rho_a$ is quadratic in $x$, but the coefficient of $x^2$ is totally different.

When $|\Delta|>\hbar J_0$, this relation can be written as:
\begin{equation}
\begin{aligned}
\rho_a=2m|c_0|^2\frac{J_0^2}{(\sqrt{|\Delta+\hbar J_0|}+\sqrt{|\Delta-\hbar J_0|})^2}x^2\\+O(x^3).
\end{aligned}
\end{equation}
If $|\Delta| \gg \hbar J_0$, our solution converges to Ref.~\cite{S25}'s result. However, for $|\Delta| < \hbar J_0$, the expression simplifies to
\begin{equation}
    \rho_a = \frac{m |c_0|^2 J_0 }{\hbar}x^2,
\end{equation}
which is independent of $\Delta$, producing a plateau in this regime. However, the model in Ref.~\cite{S25} fails to show this plateau.  
\\

\noindent{\textbf{Bohmian mechanics for quantum tunnelling}}

Since the Schrodinger equations are not in standard form, the Bohmian mechanics equations for quantum tunnelling between two coupled waveguides should be rederived.

We start from the wavefunction in the forms of:
\begin{equation}
\begin{aligned}
\psi_m=R_m e^{iS_m},\\
\psi_a=R_a e^{iS_a},
\label{wavefunction}
\end{aligned}
\end{equation}
and substitute them into Schrodinger equation. The real parts correspond to two equations:
\begin{equation}
\begin{aligned}
-\hbar\partial_t S_m =&\frac{(\hbar\partial_x S_m)^2}{2m}-\frac{\hbar^2}{2m}\frac{\partial_x^2 R_m}{R_m}+V_0\\
&+\hbar J_0 \left(\frac{R_a}{R_m}\text{cos}(S_a-S_m)-1 \right),\\
-\hbar\partial_t S_a =&\frac{(\hbar\partial_x S_a)^2}{2m}-\frac{\hbar^2}{2m}\frac{\partial_x^2 R_a}{R_a} +V_0\\
&+\hbar J_0 \left(\frac{R_m}{R_a} \text{cos}(S_m-S_a)-1 \right),
\label{HJ}
\end{aligned}
\end{equation}
which are the Hamilton-Jacobi equations of the hidden variable particles. Here, $-\hbar\partial_t S_i$ is the total energy of particles of these two modes, $\frac{(\hbar\partial_x S_i)^2}{2m}$ is the kinetic energy, $-\frac{\hbar^2}{2m}\frac{\partial_x^2 R_i}{R_i}$ is the quantum potential, $V_0$ is the external potential, and $\hbar J_0 (\frac{R_j}{R_i}\text{cos}(S_j-S_i)-1)$ is the coupling energy between these two modes.

The imagine parts also correspond to two equations:
\begin{equation}
\begin{aligned}
\partial_t(R^2_m)+\partial_x \left(\frac{R^2_m\hbar \partial_xS_m}{m} \right)+2J_0R_aR_m \text{sin}(S_m-S_a)=0,\\
\partial_t(R^2_a)+\partial_x \left(\frac{R^2_a\hbar \partial_xS_a}{m} \right)+2J_0R_mR_a \text{sin}(S_a-S_m)=0,
\label{new conti}
\end{aligned}
\end{equation}
which are the continuity equations for two coupled waveguide system. Noting that (i) another term appears in continuity equations, which describe the coupling between two waveguides; and (ii) the $x$-direction velocity of these two modes can still be written as:
\begin{equation}
\begin{aligned}
v_{x, m}&=\frac{\hbar}{m}\partial_x S_m=\frac{\hbar}{m}\text{Im}\frac{\psi^*_m\partial_x\psi_m}{\psi^*_m\psi_m},\\
v_{x, a}&=\frac{\hbar}{m}\partial_x S_a=\frac{\hbar}{m}\text{Im}\frac{\psi^*_a\partial_x\psi_a}{\psi^*_a\psi_a}.
\end{aligned}
\end{equation}
\\

\noindent{\textbf{Solving $\rho_{a,B}$ using Bohmian mechanics}}


Since the transmitted wave is monochromatic wave, the time evolution part is $e^{-i\omega t}$, so $\rho$ doesn't contain $t$, namely $\partial_t \rho = 0$. The remaining part of the continuity equation of auxiliary mode is :
\begin{equation}
\partial_x(\rho_{a,B} v_{a, x})+j_0=0,
\label{continuity reduced}
\end{equation}
where $j_0=2J_0 R_m R_a \text{sin}(S_a-S_m) = -i J_0 (\psi_a \psi^*_m -\psi^* _ a\psi_m)$.

When $x$ is sufficiently small, a Taylor expansion of $v_{a, x}$ and $j_0$ at $x=0$ leads to:
\begin{equation}
\begin{aligned}
v_{a, x}&=\frac{\hbar}{m}\text{Im}(i\frac{k_-+k_+}{2}+O(x^1)),\\
j_0&=|c_0|^2(\frac{J_0}{2}(-k_--k_-^*+k_++k_+^*)x+O(x^2)).
\end{aligned}
\end{equation}
Substituting them into the continuity equation~\ref{continuity reduced}, we have:
\begin{equation}
\begin{aligned}
&\partial_x \left[\rho_{a,B} \frac{\hbar}{m} \text{Im} \left(i\frac{k_-+k_+}{2}+O(x^1) \right) \right]\\
&+|c_0|^2\frac{J_0}{2}(-k_--k_-^*+k_++k_+^*)x+O(x^2)=0.
\end{aligned}
\end{equation}
After integration, we have:
\begin{equation}
\begin{aligned}
\rho_{a,B}&=|c_0|^2\frac{\frac{J_0}{4}(k_-+k_-^*-k_+-k_+^*)x^2+O(x^3)}{\frac{\hbar}{m} \text{Im} \left(i\frac{k_-+k_+}{2}+O(x^1) \right)},\\
&=|c_0|^2\frac{mJ_0}{\hbar}\frac{k_-+k_-^*-k_+-k_+^*}{k_-+k_-^*+k_++k_+^*}x^2+O(x^3).
\label{population reproduce}
\end{aligned}
\end{equation}

When $\Delta>\hbar J_0$, equation~\ref{population reproduce} becomes:
\begin{equation}
\begin{aligned}
\rho_{a,B}&=|c_0|^2\frac{mJ_0}{\hbar}\frac{\sqrt{\Delta+\hbar J_0}-\sqrt{\Delta-\hbar J_0}}{\sqrt{\Delta+\hbar J_0}+\sqrt{\Delta-\hbar J_0}}x^2+O(x^3)\\
&=2m|c_0|^2\frac{J_0^2}{(\sqrt{\Delta+\hbar J_0}+\sqrt{\Delta-\hbar J_0})^2}x^2+O(x^3).
\end{aligned}
\end{equation}

When$|\Delta|<\hbar J_0$, equation~\ref{population reproduce} becomes: 
\begin{equation}
\rho_a=|c_0|^2\frac{mJ_0}{\hbar}x^2+O(x^3).
\end{equation}

When$\Delta<-\hbar J_0$, we perform the analytic continuation:
\begin{equation}
\Delta=\lim_{\epsilon\rightarrow 0}\Delta+i\epsilon,
\end{equation}
where $\epsilon$ is a small value. Noting this is a common technique in quantum field theory, which involves selecting a complex part approaching zero through analytic continuation of real-valued quantities. The rationale here lies in the fact that, even in the case of evanescent waves where the average value of $v_x$ vanishes, the velocities of individual hidden variable particles conform to a distribution centered around these averages. Furthermore, this distribution must satisfy the equilibrium condition in a perturbative sense, where the infinitesimal imagine part corresponds to an infinitesimal perturbation of the propagation modes. Consequently, the particle number distribution is still required to obey the continuity equation under this velocity-dependent framework.

From $\sqrt{x+iy} = \sqrt{\frac{\sqrt{x^2 + y^2} + x}{2}} + i \text{sgn}(y)\sqrt{\frac{\sqrt{x^2+y^2}-x}{2}}$ (here we take the positive sign, corresponding to the physically reasonable value in physics, i.e., the evanescent wave solution), we have:
\begin{equation}
\begin{aligned}
&\lim_{\epsilon\rightarrow 0}(\sqrt{\Delta+\hbar J_0+i\epsilon}+(\sqrt{\Delta+\hbar J_0+i\epsilon})^*)\\
&=\lim_{\epsilon\rightarrow 0}2\sqrt{\frac{|\Delta+\hbar J_0|(1+\frac{\epsilon^2}{2|\Delta+\hbar J_0|^2}+O(\epsilon^4))+\Delta+\hbar J_0}{2}}\\
&=\lim_{\epsilon\rightarrow 0}\sqrt{\frac{\epsilon^2}{|\Delta+\hbar J_0|}+O(\epsilon^4)},
\end{aligned}
\end{equation}
and 
\begin{equation}
\begin{aligned}
&\lim_{\epsilon\rightarrow 0}(\sqrt{\Delta-\hbar J_0+i\epsilon}+(\sqrt{\Delta-\hbar J_0+i\epsilon})^*)\\
&=\lim_{\epsilon\rightarrow 0}2\sqrt{\frac{|\Delta-\hbar J_0|(1+\frac{\epsilon^2}{2|\Delta-\hbar J_0|^2}+O(\epsilon^4))+\Delta-\hbar J_0}{2}}\\
&=\lim_{\epsilon\rightarrow 0}\sqrt{\frac{\epsilon^2}{|\Delta-\hbar J_0|}+O(\epsilon^4)},
\end{aligned}
\end{equation}
so equation~\ref{population reproduce} becomes:
\begin{equation}
\begin{aligned}
\rho_a&=|c_0|^2\lim_{\epsilon\rightarrow 0}\frac{mJ_0}{\hbar}\frac{\sqrt{\frac{\epsilon^2}{|\Delta+\hbar J_0|}}-\sqrt{\frac{\epsilon^2}{|\Delta-\hbar J_0|}}}{\sqrt{\frac{\epsilon^2}{|\Delta+\hbar J_0|}}+\sqrt{\frac{\epsilon^2}{|\Delta-\hbar J_0|}}}x^2+O(x^3)\\
&=|c_0|^2\frac{mJ_0}{\hbar}\frac{\sqrt{|\Delta-\hbar J_0|}-\sqrt{|\Delta+\hbar J_0|}}{\sqrt{|\Delta-\hbar J_0|}+\sqrt{|\Delta+\hbar J_0|}}x^2+O(x^3)\\
&=2m|c_0|^2\frac{J_0^2}{(\sqrt{|\Delta+\hbar J_0|}+\sqrt{|\Delta-\hbar J_0|})^2}x^2+O(x^3).
\end{aligned}
\end{equation}

Hereto, we see that all these results are consistent with what we derived from Schr\"{o}dinger equation. There is no any distinction between Copenhagen interpretion and BM on the problem of quantum tunnelling. Any velocity defined from $\rho_a$ must be identical in both frameworks, and they must satisfy continuity equation.


\end{document}